\documentclass[10pt,conference]{IEEEtran}
\usepackage{latexsym}
\usepackage{amsmath,amssymb,amsthm}
\usepackage{psfrag,array}
\usepackage{graphicx}
\usepackage{url}

\usepackage{color,hyperref}
\definecolor{darkblue}{rgb}{0.0,0.0,0.3}
\hypersetup{colorlinks,breaklinks,
            linkcolor=darkblue,urlcolor=darkblue,
            anchorcolor=darkblue,citecolor=darkblue}

\newcommand{\Tr}{\ensuremath{^{\mathrm{T}}}}

\renewcommand{\vec}[1]{\ensuremath{\boldsymbol{#1}}}

\newtheorem{theorem}{Theorem}

\newcommand{\stxt}[1]{\ensuremath{_{\text{#1}}}}
\newcommand{\stxm}[1]{\ensuremath{_{\mathrm{#1}}}}
\newcolumntype{C}{>{$}c<{$}}
\newlength{\eqspace}
\begin{document}

\title{Achieving AWGN Channel Capacity with Sparse Graph Modulation and ``In the Air''  Coupling}

\author{ \parbox{5 in}{\centering Dmitri Truhachev\\
         Department of Computing Science\\
         University of Alberta, Canada\\
         {\tt\small dmitryt@ualberta.ca}}}

\maketitle

\begin{abstract}
Communication over a multiple access channel is considered. Each user modulates his signal as a superposition of redundant data streams where interconnection of data bits can be represented by means of a sparse graph. The receiver observes a signal resulting from the coupling of the sparse modulation graphs. Iterative interference cancellation decoding is analyzed. It is proved that spatial graph coupling allows to achieve the AWGN channel capacity with equal power transmissions.
\end{abstract}

\section{Introduction}

The effect of spatial graph coupling, first discovered for low-density error correction codes, attracts increasing interest from the communications research community. Iterative decoding of block low-density parity-check (LDPC) codes, despite its known strength, fails to achieve the same limits as the optimum maximum likelihood (ML) decoding. It has been shown, however~\cite{lscz10}, and then rigorously proved~\cite{kru10}, that iterative decoding thresholds of convolutional (or spatially coupled) LDPC codes coincide with the ML decoding thresholds. A coupled code is constructed by copying the protograph of an initial block code a number of times and then connecting the neighboring copies to form a chain. The graphs located at the ends of the chain contain variable nodes with fewer constraints, since these graphs are connected to their neighbors from only one side. As a result, iterative decoding progress initiates at the ends of the chain, due to the effect of the slight irregularity, and then propagates through the entire chain. The principle of spatial graph coupling has proven to be applicable to several other areas, including multiple user detection, compressive sensing, and quantum coding.


In this paper, we focus our attention on spatial graph coupling in relation to multiple access communications. The capacity and the achievable rate region for the multiple access channel (MAC) are well known. The corner points of the rate region polytope can be achieved by successive (onion peeling) decoding~\cite{CovTho06} while the middle part is achieved by rate splitting and/or time sharing added to the onion peeling~\cite{RimUrb01}. However, communication approaches that allow for more robust and less complex joint parallel detection/decoding, such as code-division multiple access (CDMA), have had limited success in achieving the inner points of the MAC rate region. The equal power user case is typically the most difficult, since all users happen to be operating under same conditions, and there is no structural irregularity to initiate the decoding convergence. It has been shown that regular random CDMA can only support a fixed system load\footnote{The system load $\alpha$ is defined as the ratio of the number of supported users $K$ to the number of available signal dimensions $N$ (or chips in CDMA), i.e, $\alpha=K/N$.} equal to $\alpha= 1.49$~\cite{Tanaka2002} with equal power users, while sparse synchronous CDMA can only support $\alpha=2.07$~\cite{TruSchKrz2007_IT}.

In this work we consider a multiple access multiuser situation where each user modulates his signal as a sum of redundant, independent, equal power, data streams, as described in \cite{TruSchKrz2007_IT}\cite{GM2010}. The modulated signal allows for sparse graph representation similar to the LDPC code's Tanner graph. The sparse graphs of the transmitted signals couple together ``in the air,'' and the receiver observes a coupled sequence of graphs which is later processed by an iterative detector/decoder. We focus on a communication scenario in which all users transmit an equal number of equal power BPSK data streams that are encoded by error correction codes optimal for the binary input additive white Gaussian noise (AWGN) channel. We then prove that the (real-valued) AWGN channel capacity can be achieved exactly by such a system.

\section{System Model}

We consider the generalized modulation type format~\cite{GM2010}. Each user modulates a number of independent data streams and transmits their sum. One data stream is modulated as follows. First the data is encoded by a binary error control encoder to produce a binary data stream $\{u_{l}\}^L_{l=1}$. The stream is first duplicated $M$ times resulting in $M$ identical data sub-streams, $\{u_{m,l}\}^{L}_{l=1}$, $m=1,2,\ldots,M$. Each sub-stream $\{u_{m,l}\}^{L}_{l=1}$ is permuted by an interleaver $\pi_{m}$ to produce $\{\tilde{u}_{m,l}\}^{L}_{l=1}$, $m \in \{1,2,\ldots,M\}$. At the next step each bit of $\{\tilde{u}_{m,l}\}^{L}_{l=1}$ is multiplied by an $N$-dimensional signature vector $\vec{s}_{m}$. Signature vector maps the data into the available resource space. This operation is similar to spreading in a CDMA system.
Each of the $M$ sub-streams is multiplied by an amplitude $\sqrt{P/M}$ and they are all added up to produce one modulated signal stream
\begin{equation}
\vec{v}_{l} = \sum_{m=1}^{M} \sqrt{\frac{P}{M}} \tilde{u}_{m,l} \vec{s}_{m} \quad l=1,2,\ldots,L\ .
\label{eq:userkTx}
\end{equation}
Each signature sequence $\vec{s}_{m} = \{s_{m,1},s_{m,2},\ldots,s_{m,N}\}$ is real and energy normalized such that $\mathbb{E} |s_{m,i}|^2 = \frac{1}{N}$, $i=1,2,\ldots,N$.
We also require~\footnote{The following property can be easily guaranteed by
choosing $s_{m,i} \in \left\{-\frac{1}{\sqrt{N}},\frac{1}{\sqrt{N}}\right\}$ iid Bernoulli with probability $1/2$. However, other signature sequence choices are also possible.} $\mathbb{E} \vec{s}_{m_1} \vec{s}\Tr_{m_2}  = 1/N$, $m_1 \neq m_2$.

\begin{figure}[t]
\setlength{\unitlength}{1mm}
   \begin{picture}(85,37)
   \put(0,-2){\includegraphics{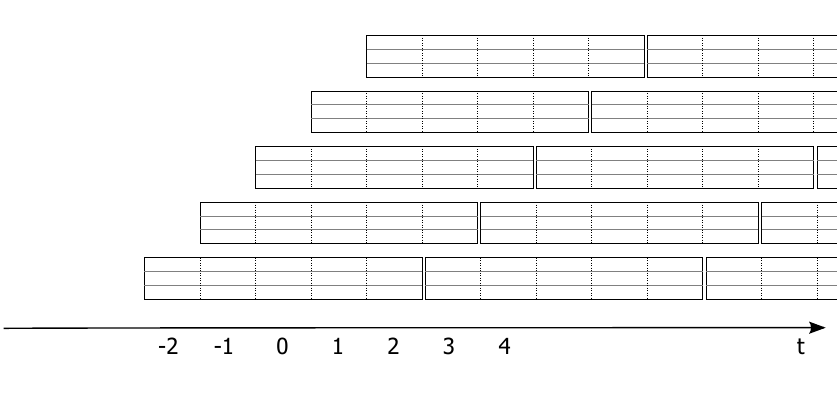}
   }
\end{picture}
\caption{Transmitted packets coupling in the channel, $W=2$.}
\label{Fig:Packets}
\end{figure}

We consider the case in which $2W+1$ users transmit $K/(2W+1)$ data streams each, producing total system load of $\alpha=K/N$. The power of each modulated stream is $P=1$ and communication happens over the real-valued AWGN channel with noise power $\sigma^2$. We focus on a scenario in which the packets of the users are received asynchronously as shown in Fig.~\ref{Fig:Packets}. Each packet is divided into $2W+1$ sections of length $(2W+1)/L$ bits each. At time slot $t=-W$ the first user starts to transmit his packet and transmits one section. At time $t=-W+1$ the second user joins and transmits one section while the first user transmits his second section. At time $t=-W+2$ the third user joins and so on. Once a user finishes transmitting his packet he immediately starts to transmit his next packet. 

We consider two types of receivers. The first receiver is a {\em two-stage receiver} where the first stage is an iterative parallel interference cancellation (PIC) which layers the received data streams and the second stage is an error control decoding performed for all layered streams independently and in parallel. The first, PIC stage, can be described as follows. At every iteration the received signals are filtered using the signature sequences $\vec{s}_{k,m}$. Assuming a simple matched filter, the filtered signal of data bit $u_{k,l}$ from data stream $k \in \{1,2,\ldots,K\}$, and sub-stream $m \in \{1,2,\ldots,M\}$ equals
\begin{align}
q_{k,m,l} &= \frac{1}{\sqrt{M}} \vec{s}\Tr_{k,m} \bigl( u_{k,m,l} \vec{s}_{k,m} + \hspace*{-4mm}\sum_{(k_1,m_1) \neq (k,m)} \tilde{u}_{k_1,m_1,l} \vec{s}_{k_1,m_1} \bigr)  \nonumber \\
          &= \frac{1}{\sqrt{M}} u_{k,m,l} + I_{k,m,l}
\end{align}
where $I_{k,m,t}$ is the interference, suppressed by the filtering. Each bit $u_{k,l}$ has $M$ replicas $u_{k,m,l}$ which are dispersed along the packet. The power $x_m$ of the interference term $I_{k,m,l}$ depends on the section of the packet where replica $u_{k,m,l}$ belongs to. The signal power of each replica is $1/M$. We estimate the transmitted bits by summing up weighted received replicas $q_{k,m,l}$ and applying the conditional expectation estimate
\begin{equation}
\hat{u}_{k,m,l} = \tanh \left(\sum_{m_1 \neq m} \xi_{m_1} q_{k,m_1,l} \right)\ .
\label{eq:bitest}
\end{equation}
The optimal weighting coefficients
\[
\xi_m = \frac{1}{x_m}\left(\sum_{m'=1}^M \frac{1}{x_{m'}} \right)^{-1}
\]
ensure that the signal-to-noise-and-interference ratio (SINR) of the weighted sum inside the parenthesis in (\ref{eq:bitest}) is maximized and equals
\begin{equation}
z_1 =  \frac{1}{M} \sum_{m_1 \neq m} \frac{1}{x_{m_1}}\ .
\label{eq:SINR}
\end{equation}
The bit estimates (\ref{eq:bitest}) are used to reconstruct the entire transmitted signals (\ref{eq:userkTx}) and perform interference cancellation by subtraction of the reconstructed signals from the received signal. The iterative process is repeated $i$ times until the SINR $z_i$ of the data bits at iteration $i$ becomes greater than the decoding threshold $\theta$ of the error control code used for their encoding. The error control decoding (second stage) can than be applied to all the packets (received at given time) for the final error correction.

The second receiver, which we call a {\em modified successive interference cancellation (SIC) receiver} is working as follows. It starts with the first stage of the two-stage receiver described above. However, the error correction decoding of the data streams for which $z_i > \theta$ is performed after each interference cancellation iteration. These decoded streams are then removed from the interference pool. The modified SIC receiver is slightly more complex since it requires a feedback loop between the decoder and the interference canceller. However, the modified SIC allows for simpler analysis.

\section{Analysis}

We start with an analysis of the iterative interference cancellation, i.e., the first stage of the two-stage receiver. We track the evolution of the noise-and-interference variance which we denote by $x_i^t$, and the SINR of the data bits which we denote by $z_i^t$. The upper index $t$ indicates the time slot of interest and $i$ is the iteration number. A discrete system of equations describing the evolution of $x_i^t$ and $z_i^t$ have been derived in~\cite{SchTruISIT11}
\begin{align}
x_i^t &= \frac{\alpha}{2W+1} \sum_{j=-W}^{W} g\left(z_i^{t+j}\right) + \sigma^2  &i\geq 0, t \geq 1\label{eq:xdis} \\
z_i^t &= \frac{1}{2W+1}\sum_{j=-W}^{W} \frac{1}{x_{i-1}^{t+j}}  &i> 0, t \geq W+1 \label{eq:zdis}
\end{align}
where the function $g(\cdot)$ given by
\[
g(a) = \mathbb{E}\left[\left(1-\tanh\left(a+\xi\sqrt{a}\right)\right)^2\right]\ , \quad\xi \sim \mathcal{N}(0,1)\ ,
\]
determines the mean squared error of the data bits with SINR $a$ and $\mathcal{N}(0,1)$ denotes a standard normal random variable. Function $g(\cdot)$ is a continuous differentiable function that is strictly decreasing from $g(0)=1$ to $g(+\infty)=0$.
Variable $z_i^t$ defines the SINR of the packet starting at time $t-W$ and finishing at $t+W$, i.e., centered at time $t$.
SINR $z_i^t$ is computed from the SINRs of the $M$ partitions uniformly distributed along
the packet. The SINRs $\frac{1}{x_{i-1}^{t+l}}$ of the partitions are averaged in (\ref{eq:zdis}), just like in (\ref{eq:SINR}). Packets that start between time $t-2W$ and $t$ contribute to the noise-and-interference variance at time $t$ in (\ref{eq:xdis}).


The iterative system is initialized by
\begin{equation}
z_0^t = \left\{ \begin{array}{ll}
             +\infty, & \quad t \leq W \\
             0, & \quad t \geq W+1\ ,
           \end{array} \right.
\label{eq:zdis_ini}
\end{equation}
since  the packets that start before time $t=0$ are completely known (or absent) and, therefore, their SINR is $+\infty$. These packets are centered at time before $W+1$. The packets
centered at time $t=W+1$ and after are completely unknown. Hence, their SINR is $0$.



Considering $W \rightarrow \infty$ and normalizing the packet length to $1$ the equations (\ref{eq:xdis}) and (\ref{eq:zdis})
can be transformed to
\begin{align}
x_i^t &= \alpha \int_{-1/2}^{1/2}  g(z_i^{t+\tau}) d \tau + \sigma^2 &t \geq -\frac{1}{2}, i \geq 0 \label{eq:xcon} \\
z_i^t &= \int_{-1/2}^{1/2} \frac{1}{x_{i-1}^{t+\tau}} d \tau &t \geq 0, i > 0 \label{eq:zcon}
\end{align}
with a shift of the time origin from $1/2$ to $0$  which is done for notation simplicity.
The initialization is done using a step function
\begin{equation}
z_0^t = \left\{ \begin{array}{ll}
             +\infty, & \quad t < 0 \\
             0, & \quad t \geq 0\ .
           \end{array} \right.
\label{eq:zcon_ini}
\end{equation}

The SINR of the data bits at iteration $I$ of the interference cancellation stage equals $z_I^t$. The second, error correction decoding stage, of the two-stage
reception is completed successfully iff $z_I^t > \theta$ where $\theta$ is the error correction code threshold.

Contrary to the two-stage receiver, the modified SIC receiver is performing decoding after every interference cancellation iteration. SIC can be described by (\ref{eq:xcon}) together with
\begin{equation}
z_{\mathrm{SIC},i}^t = \left\{ \begin{array}{ll}
             +\infty, & \quad \int_{-1/2}^{1/2} \frac{1}{x_{i-1}^{t+\tau}} d \tau > \theta \\
             \int_{-1/2}^{1/2} \frac{1}{x_{i-1}^{t+\tau}} d \tau, & \quad \int_{-1/2}^{1/2} \frac{1}{x_{i-1}^{t+\tau}} d \tau \leq \theta\ .
           \end{array} \right.
\label{eq:zconSIC}
\end{equation}
replacing equation (\ref{eq:zcon}). We notice that the operator $F(\cdot)$ defining the evolution of the SINR throughout the iterations
via (\ref{eq:xcon}) and (\ref{eq:zconSIC}),
\begin{equation}
z_{i+1}^t = F(z_i^t)\ ,
\label{eq:F}
\end{equation}
is monotone, i.e., if $z_1(t) \leq z_2(t)$ for $t \in (-\infty,+\infty)$, then $F(z_1(t)) \leq F(z_2(t))$ for $t \in (-\infty,+\infty)$.


\begin{theorem}
\label{thm:speff}
Spectral efficiency of
\begin{equation}
\alpha \mathcal{C}_{\mathrm{BIAWGN}}\left(\frac{1}{\alpha} \ln \frac{\alpha+\sigma^2}{\sigma^2}-\epsilon\right)
\label{eq:siccapeq}
\end{equation}
is achievable by the modified SIC receiver for any $\alpha \in [0,\infty)$ and $\epsilon>0$.
\end{theorem}
\begin{proof}
We proceed with (\ref{eq:zcon_ini}) and (\ref{eq:xcon}) calculating
\begin{align}
x_0^t &=  \int\limits_{t-1/2}^{t+1/2} \alpha g\left(z_0^{\tau}\right) d \tau + \sigma^2 = \left\{ \begin{array}{ll}
             \alpha t+\frac{\alpha}{2} + \sigma^2, &t \in \left[-\frac{1}{2},\frac{1}{2}\right] \\
             \alpha + \sigma^2, &t \in \left(\frac{1}{2},\infty\right)
           \end{array} \right. \nonumber
\end{align}
that after the application of (\ref{eq:zcon}) implies
\begin{align}
z_1^0 &=  \int\limits_{-1/2}^{1/2} \frac{1}{x_{0}^{\tau}} d \tau  =  \int\limits_{-1/2}^{1/2} \frac{1}{\alpha \left(\tau+\frac{1}{2}\right) + \sigma^2} d \tau  = \frac{1}{\alpha} \ln \frac{\alpha+\sigma^2}{\sigma^2}, \nonumber\\
z_1^\tau &= \frac{1}{\alpha} \ln \frac{\alpha+\sigma^2}{\sigma^2} - \frac{\alpha}{\sigma^2(\alpha+\sigma^2)}\delta+ o(\delta), \quad \tau \in [0,\delta]
\end{align}
for small $\delta >0$. 
Let us assume that the binary error correction codes for all data streams are of rate
\begin{equation}
\mathcal{C}\stxt{BIAWGN}\left(\frac{1}{\alpha} \ln \frac{\alpha+\sigma^2}{\sigma^2} - \frac{\alpha}{\sigma^2(\alpha+\sigma^2)}\delta+ o(\delta) \right),
\label{eq:rate}
\end{equation}
i.e., these codes have threshold
\[
\theta = \frac{1}{\alpha} \ln \frac{\alpha+\sigma^2}{\sigma^2} - \frac{\alpha}{\sigma^2(\alpha+\sigma^2)}\delta+ o(\delta)
\]
In this case (\ref{eq:zconSIC}) implies
\[
z_{\mathrm{SIC},1}^t = +\infty, \quad t < \delta\ ,
\]
indicating that the packets centered at $t \in [0,\delta)$ are successfully decoded and cancelled.
The situation will repeat at the next iteration since $z_{\mathrm{SIC},1}^t \geq z_0^{t-\delta}$, $t\in(-\infty,+\infty)$ implies $z_{\mathrm{SIC},2}^t = +\infty$ for $t \in [\delta, 2\delta)$
from the monotonicity of the SINR evolution operator (\ref{eq:F}). Thus, the convergence of the SINR to $+\infty$ will continue at the speed of at least $\delta$ per iteration. The total achievable sum rate is (see (\ref{eq:rate}))
\[
\alpha \left( \mathcal{C}\stxt{BIAWGN} \left(\frac{1}{\alpha} \ln \frac{\alpha+\sigma^2}{\sigma^2} - \frac{\alpha}{\sigma^2(\alpha+\sigma^2)}\delta+ o(\delta) \right)\right)\ .
\]
By choosing appropriately small $\delta$ we obtain the statement of the Theorem.
\end{proof}

%

Let us define the total system SNR parameter $s= \frac{\alpha}{\sigma^2}$. The next theorem states that for any fixed $s$ the AWGN channel capacity is achievable by the
modified SIC decoder. Let us consider $\sigma^2 = \alpha/s$, i.e., we are keeping $s$ fixed.  We denote the limiting spectral efficiency of the system considered in Theorem~\ref{thm:speff}
and the corresponding SNR per bit by
\begin{align}
\mathcal{C}\stxt{eff}(\alpha,s) &= \alpha \mathcal{C}_{\mathrm{BIAWGN}}\left(\frac{1}{\alpha} \ln (1+s)\right) \label{eq:ceffalphs}\\
\frac{E\stxt{b}}{N_0}(\alpha,s) &= \frac{1}{2 \mathcal{C}_{\mathrm{BIAWGN}}\left(\frac{1}{\alpha} \ln (1+s)\right) \sigma^2}\ . \label{eq:snralphs}
\end{align}
We also define the real-valued AWGN channel capacity $\mathcal{C}\stxt{AWGN}\left( \frac{E\stxt{b}}{N_0} \right)$ for given $\frac{E\stxt{b}}{N_0}$ as the root of the equation
\begin{equation}
\mathcal{C}\stxt{AWGN}\left( \frac{E\stxt{b}}{N_0} \right) = \frac{1}{2} \log_2\left(1 + 2\mathcal{C}\stxt{AWGN}\left( \frac{E\stxt{b}}{N_0} \right) \frac{E\stxt{b}}{N_0}\right)\ .
\label{eq:cawgn}
\end{equation}

Finally, we state the main result of the paper.
\begin{theorem}
\label{thm:capach}
\[
\lim_{\alpha \rightarrow \infty} \mathcal{C}\stxm{eff}(\alpha,s) = \lim_{\alpha \rightarrow \infty} \mathcal{C}\stxm{AWGN}\left( \frac{E\stxm{b}}{N_0}(\alpha,s) \right) = \frac{1}{2}\log_2 (1+s)
\]
\end{theorem}
\begin{proof}
The capacity of the binary input AWGN channel for SNRs $\gamma<1$ can be bounded as follows~\cite{Gal06}
\begin{equation}
\frac{1}{2\ln 2} \gamma + A_1 \gamma^2 \leq \mathcal{C}_{\mathrm{BIAWGN}}(\gamma) \leq \frac{1}{2\ln 2} \gamma + A_2 \gamma^2 \label{eq:Galbnd}
\end{equation}
where $A_1 < A_2$ are constants. Therefore, application of (\ref{eq:Galbnd}) to (\ref{eq:ceffalphs}) implies
\begin{multline}
\frac{1}{2} \log_2 (1+s) +  A_1 \frac{\ln^2 (1+s)}{\alpha} \leq \mathcal{C}\stxt{eff}(\alpha,s) \\ \leq \frac{1}{2} \log_2 (1+s) +  A_2 \frac{\ln^2 (1+s)}{\alpha}
\end{multline}
while application of (\ref{eq:Galbnd}) to (\ref{eq:snralphs}) implies
\begin{multline}
\frac{s}{ \log_2 (1+s) +  2 A_2 \frac{\ln^2 (1+s)}{\alpha}} \leq \frac{E\stxt{b}}{N_0}(\alpha,s) \\ \leq \frac{s}{ \log_2 (1+s) +  2 A_1 \frac{\ln^2 (1+s)}{\alpha}}\ .
\end{multline}
Taking the limit in the above inequalities we obtain
\begin{align}
&\lim_{\alpha \rightarrow \infty} \mathcal{C}\stxt{eff}(\alpha,s) =  \frac{1}{2}\log_2 (1+s) \label{eq:ceffs}\\
&\lim_{\alpha \rightarrow \infty} \frac{E\stxt{b}}{N_0}(\alpha,s) =  \frac{s}{ \log_2 (1+s)} \label{eq:ebn0s}
\end{align}
Equations~\ref{eq:ceffs} and \ref{eq:ebn0s} imply equality
\[
\lim_{\alpha \rightarrow \infty}  \frac{E\stxt{b}}{N_0}(\alpha,s)  = \frac{2^{2\lim_{\alpha \rightarrow \infty} \mathcal{C}\stxt{eff}(\alpha,s)}-1}{2\lim_{\alpha \rightarrow \infty} \mathcal{C}\stxt{eff}(\alpha,s)}
\]
which ensures (see (\ref{eq:cawgn}))
\[
\lim_{\alpha \rightarrow \infty} \mathcal{C}\stxm{eff}(\alpha,s) = \lim_{\alpha \rightarrow \infty} \mathcal{C}\stxm{AWGN}\left( \frac{E\stxm{b}}{N_0}(\alpha,s) \right)\ .
\]
The Theorem is proved.
\end{proof}

\begin{figure}[t]
\setlength{\unitlength}{1mm}
   \begin{picture}(90,56)
   \put(0,-3){\includegraphics{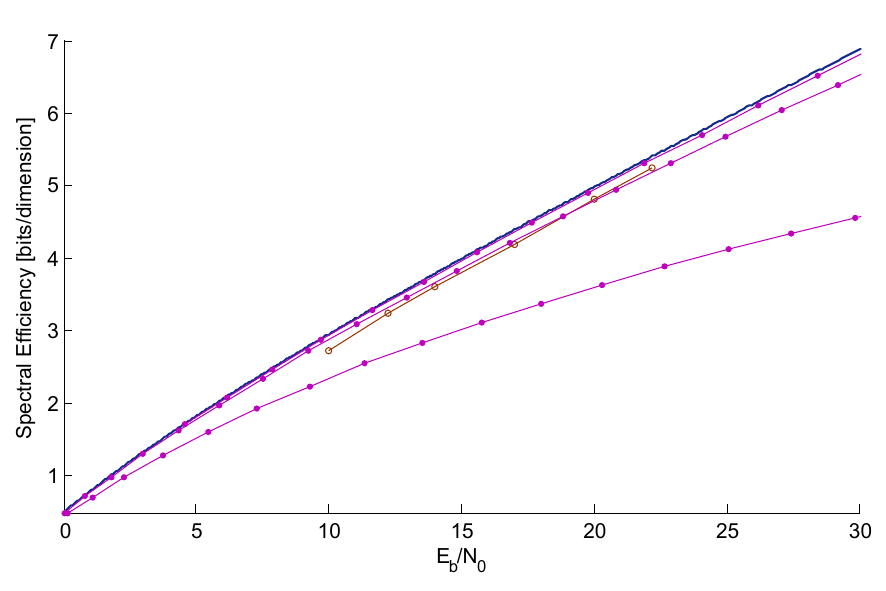}}
\end{picture}
\caption{Achievable spectral efficiency for the modified SIC receiver (magenta), two-stage receiver (brown) and the AWGN channel capacity (blue).}
\label{Fig:CapCeff}
\end{figure}

Spectral efficiency $\mathcal{C}\stxm{eff}(\alpha,s)$ achieved by the modified SIC receiver (see Theorem~\ref{thm:speff}) is plotted in Fig.~\ref{Fig:CapCeff}. The three magenta curves correspond to $\alpha =10,100,$ and $500$ (from bottom to top). For each curve $\alpha$ is kept constant while $s$ is varying. The channel capacity $\mathcal{C}\stxm{AWGN}$ is given by the blue curve. Finally, the brown curve plots spectral efficiency achieved by the two-stage receiver.

\pagebreak

\vspace*{-7mm}

\section{Conclusions}
In this paper, it has been proven that sparse graph modulation with spatial coupling can achieve AWGN channel capacity under modified SIC reception. The sparse graph modulation format is based on superposition of low-rate redundant data streams which can be easily designed to operate near binary input AWGN channel capacity. Numerical results also demonstrate that the two-stage reception where parallel interference cancellation is followed by parallel error control decoding is capable of operating within 1dB of the channel capacity.



\begin{thebibliography}{99}
\small


\bibitem{lscz10}
M.~Lentmaier, A.~Sridharan, D.~J. {Costello, Jr.}, and {K. Sh. Zigangirov},
  ``Iterative decoding threshold analysis for {LDPC} convolutional codes,''
  \emph{IEEE Trans. Inf. Theory}, vol.~56, no.~10, pp.
  5274--5289, Oct. 2010.

\bibitem{kru10}
S.~Kudekar, T.~Richardson, and R.~Urbanke, ``Threshold saturation via spatial
  coupling: why convolutional {LDPC} ensembles perform so well over the
  {BEC},'' in \emph{Proc. IEEE Int. Symp. on Inf. Theory},
  Austin, TX, June 2010.

\bibitem{SchTruISIT11}
C.~Schlegel and D.~Truhachev, ``Multiple Access Demodulation in the Lifted Signal
Graph with Spatial Coupling,'' in \emph{Proc. IEEE Int.
  Symp. on Inf. Theory}, St. Petersburg, Russia, Aug. 2011.

%
%

\bibitem{CovTho06} T.~Cover and J.~Thomas, {\em Elements of Information Theory}, Wiley \& Sons, New York, 2006.

\bibitem{RimUrb01} B.~Rimoldi and R.~Urbanke, ``A Rate Splitting Approach to the Gaussian Multiple Acess Channel''
{\em IEEE Trans. on Inf. Theory}, vol. 42, no. 2, pp. 364--375, Mar. 1996.

\bibitem{TruSchKrz2007_IT} D.~Truhachev, C.~Schlegel and L.~Krzymien,
``A Two-Stage Capacity Achieving Demodulation/Decoding Method for
Random Matrix Channels'', {\em IEEE Trans. on Inf. Theory},
vol. 55, pp. 136--146, Jan. 2009.

\bibitem{GM2010} C.~Schlegel, M.~Burnashev, and D.~Truhachev, ``Generalized Superposition Modulation and Iterative Demodulation:
A Capacity Investigation,'' {\em Hindawi Journal of Electrical and Computer Engineering}, vol. 2010, Article ID 153540, Sep. 2010.

\bibitem{Tanaka2002} T.~Tanaka ``A statistical mechanics approach to large-system
analysis of CDMA multiuser detectors'', {\em IEEE Trans. Inf.
Theory}, vol.~48, no.~11, pp.~2888--2910, Nov. 2002.

\bibitem{Gal06} R.~Gallager, {\em Information Theory and Reliable Communications}, Wiley \& Sons, New York, 1968.
\end{thebibliography}
\end{document}